\documentclass[journal=jacsat,manuscript=article]{achemso}
\usepackage{chemformula} 
\usepackage[T1]{fontenc} 
\usepackage{epsfig}
\usepackage{graphicx}
\usepackage{titlesec}
\usepackage{dcolumn}
\usepackage{amsthm,amsmath}
\usepackage{comment}
\usepackage[colorlinks]{hyperref}
\usepackage{xcolor}
\usepackage[T1]{fontenc}
\usepackage{mathrsfs}
\usepackage{longtable}
\usepackage{orcidlink}
\usepackage{amsmath}
\usepackage{tabularx}
\usepackage{tabularray}
\usepackage{threeparttable}
\usepackage{caption}
\usepackage[utf8]{inputenc}
\usepackage[T1]{fontenc}
\usepackage[normalem]{ulem}

\usepackage{resizegather}
\UseRawInputEncoding

\author{Somesh Chamoli}
\affiliation[Unknown University]
{\small Department of Chemistry, Indian Institute of Technology Bombay, Powai, Mumbai 400076, India}

\author{Malaya K. Nayak}
\affiliation[Unknown University]
{\small Theoretical Chemistry Section, Bhabha Atomic Research Centre, Trombay,
Mumbai 400085, India}
\alsoaffiliation{\small Homi Bhabha National Institute, BARC Training School Complex, Anushakti Nagar, Mumbai 400094, India}
\author{Achintya Kumar Dutta}
\email{achintya@chem.iitb.ac.in}
\affiliation[Unknown University]
{\small Department of Chemistry, Indian Institute of Technology Bombay, Powai, Mumbai 400076, India}
\alsoaffiliation{\small Department of Inorganic Chemistry, Faculty of Natural Sciences, Comenius University, Ilkovi\v{c}ova 6, Mlynsk\'a dolina 84215 Bratislava, Slovakia}

\title[An \textsf{achemso} demo]
  {\Large A Reduced Cost Two-component Relativistic Equation-of-Motion Coupled Cluster Method for Ionization Potential}

\keywords{spinors, \LaTeX}

\begin{document}

\maketitle
\begin{abstract}
 We report an efficient implementation of the ionization potential (IP) variant of the equation-of-motion coupled cluster (IP-EOM-CC) method based on the exact two-component atomic mean field (X2CAMF) framework, utilizing Cholesky decomposition (CD) and frozen natural spinors (FNS). The CD approximation significantly reduces memory demands, whereas the FNS approximation lowers the number of floating-point operations. Together, these techniques make the method computationally efficient for accurate relativistic IP-EOM-CC calculations of molecules containing heavy elements. The calculated IP values are almost identical to those obtained by the four-component relativistic IP-EOM-CC method. Benchmark studies show good agreement with experimental ionization energies and photoelectron spectra, demonstrating the method's reliability.  The practical applicability of the approach is demonstrated by IP calculations on the medium-sized [I(H$_{2}$O)$_{12}$]$^{-}$ complex, with 1698 virtual spinors. 
\end{abstract}

\section{Introduction}
Theoretical simulations are essential for interpreting ionization-induced phenomena observed in experiments. A key aspect of it is the precise determination of ionization energies (IEs), which have implications in molecular electronic structure, spectroscopy, and chemical reactivity. In systems containing heavy elements, accurately simulating atomic and molecular spectroscopy requires a balanced consideration of both relativistic and electron correlation effects to ensure reliable predictions. The equation-of-motion coupled cluster (EOM-CC) method\cite{RevModPhys.40.153,10.1063/1.464746,10.1063/1.468592} has emerged as a powerful tool for computing ionization energies due to its robust treatment of electron correlation and its ability to describe multiple ionized states in a single calculation. However, incorporating relativistic effects into EOM-CC calculations significantly increases computational cost, particularly in Dirac-Coulomb Hamiltonian-based four-component relativistic methods that explicitly account for spin-orbit coupling. To achieve a balance between accuracy and computational efficiency, two-component relativistic approaches\cite{PhysRevA.33.3742,https://doi.org/10.1002/(SICI)1097-461X(1996)57:3<281::AID-QUA2>3.0.CO;2-U,10.1063/1.473860,NAKAJIMA1999383,BARYSZ2001181,10.1063/1.3159445,saue:hal-00662643} provide a promising alternative. These methods can effectively capture major relativistic effects while substantially reducing computational cost compared to their four-component counterparts. Among the various two-component approaches, the exact two-component theory (X2C)\cite{10.1063/1.473860,10.1063/1.3159445,10.1063/1.2137315,10.1063/1.2436882,10.1093/oso/9780195140866.001.0001} has emerged as a popular alternative to incorporate relativistic effects into electronic structure calculations. Among its multiple variations available, the X2C scheme utilizing atomic mean field (AMF)\cite{HE1996365} spin-orbit (SO) integrals, known as X2CAMF\cite{10.1063/1.5023750,doi:10.1021/acs.jpca.2c02181,10.1063/5.0095112} method stands out as a particularly effective and efficient approach. The X2CAMF method provides a balanced and accurate treatment of both spin-orbit and scalar-relativistic effects while significantly improving computational efficiency. This makes it especially well-suited for studying large systems containing heavy elements, where a precise yet cost-effective relativistic treatment is essential. The X2CAMF-based EOM-CC approach\cite{10.1063/1.5081715}, as well as the analytical gradients for EOM-CC methods within the X2CAMF framework\cite{10.1063/5.0175041}, have already been established in the literature. Although X2CAMF-based EOM-CC calculations offer a computational advantage over the four-component method, their application to medium- and large-sized molecules remains constrained by the substantial storage requirement for the two-electron integrals and intermediates involved. Additionally, the inherently high computational cost of the coupled cluster method further limits its applicability to small systems and/or moderate basis sets. This is particularly problematic for relativistic calculations, where the complex numbers are required to store the integrals, and the lack of spin symmetry makes a relativistic coupled cluster (at the singles doubles approximation) calculation at least 32 times more expensive than the corresponding non-relativistic calculations\cite{10.1093/oso/9780195140866.001.0001}.
To reduce the storage requirement of relativistic calculations, various techniques have been introduced focusing on the decomposition of two-electron integrals, including Cholesky decomposition (CD)\cite{HELMICHPARIS201938,10.1063/5.0161871,doi:10.1021/acs.jpca.4c04353} and density fitting (DF)\cite{10.1063/1.4807612,10.1063/1.4906344}. Among these, CD is often preferred, as it allows for better control over the errors compared to DF, and one does not need a preoptimized auxiliary basis set. A CD-based X2CAMF-CC and X2CAMF-EOM-CC method has been established to overcome storage constraints, allowing for calculations on medium-sized molecules\cite{doi:10.1021/acs.jctc.3c01236}. Additionally, perturbative truncation of cluster amplitudes has been explored as a means to lower the intrinsic computational expense of relativistic EOM-CC methods\cite{PhysRevA.90.062501,10.1063/1.4964859}. The use of DF or CD approximation for the integrals does not reduce the formal scaling of the integrals.  Moreover, in relativistic calculations, one often needs to uncontract the basis set to fit the small-component, which leads to a large size of the virtual space in the correlation calculations. Therefore, an effective approach to reduce the computational cost of relativistic CC calculations is to minimize floating-point operations using natural spinors\cite{10.1063/5.0085932,10.1063/5.0087243,doi:https://doi.org/10.1002/9781394217656.ch5}. Implementation of frozen natural spinors (FNS) in the four-component framework is already available in the literature to lower the computational demands of relativistic EOM-CC calculations for ionization potential\cite{10.1063/5.0125868}. However, due to the large size of the molecular integrals, application of the FNS-based four-component relativistic IP-EOM-CCSD method is restricted to small molecules only. More recently, an FNS- and CD-based X2CAMF-CC method was introduced to reduce both storage requirements and floating-point operations simultaneously\cite{doi:10.1021/acs.jctc.5c00199} for ground state coupled cluster calculations.

This manuscript aims to extend the FNS framework and CD-X2CAMF scheme to the ionization potential variant of the EOM-CC method (X2CAMF-IP-EOM-CC), enabling accurate calculations of ionization energies for medium-sized molecules containing heavy elements.

\section{Theory}
\subsection{Exact Two-Component Hamiltonian with Atomic Mean Field Integrals (the X2CAMF scheme)}
The four-component Dirac-Coulomb (DC) Hamiltonian is expressed in terms of the occupation number representation as
\begin{equation}
\label{eq:1}
\hat{H^{\text{4c}}}=\sum_{pq}{h_{pq}^{\text{4c}}}a_{p}^{\dagger}a_{q}+\frac{1}{4}\sum_{pqrs}{g_{pqrs}^{\text{4c}}}a_{p}^{\dagger}a_{q}^{\dagger}a_{s}a_{r}
\end{equation}
Here the indices \textit{p}, \textit{q}, \textit{r}, \textit{s} represent positive-energy four-component spinors within the framework of the no-pair approximation\cite{PhysRevA.22.348}.
Under the spin-separation scheme\cite{10.1063/1.466508}, the two-electron interaction matrix elements can be partitioned into spin-free (SF) and spin-dependent (SD) components when expressed in a kinetically balanced basis set\cite{10.1063/1.447865}.
\begin{equation}
\label{eq:2}
g_{pqrs}^{\text{4c}}=g_{pqrs}^{\text{4c,SF}}+g_{pqrs}^{\text{4c,SD}}
\end{equation}
By substituting Eq. (\ref{eq:2}) into Eq. (\ref{eq:1}), the four-component DC Hamiltonian can be rewritten as 
\begin{equation}
\label{eq:3}
\hat{H^{\text{4c}}}=\sum_{pq}{h_{pq}^{\text{4c}}}a_{p}^{\dagger}a_{q}+\frac{1}{4}\sum_{pqrs}{g_{pqrs}^{\text{4c,SD}}}a_{p}^{\dagger}a_{q}^{\dagger}a_{s}a_{r}+\frac{1}{4}\sum_{pqrs}{g_{pqrs}^{\text{4c,SF}}}a_{p}^{\dagger}a_{q}^{\dagger}a_{s}a_{r}
\end{equation}
Taking advantage of the local nature of the spin-orbit interaction, the SD component of the four-component DC Hamiltonian can be modeled using the atomic mean-field (AMF) approximation\cite{HE1996365}.
\begin{equation}
\label{eq:4}
\frac{1}{4}\sum_{pqrs}{g_{pqrs}^{\text{4c,SD}}}a_{p}^{\dagger}a_{q}^{\dagger}a_{s}a_{r}\approx\sum_{pq}{g_{pq}^{\text{4c,AMF}}}a_{p}^{\dagger}a_{q}=\sum_{pq}\sum_{A}\sum_{i}{n_{i,A}\hspace{0.1cm}g_{pi_{A}qi_{A}}^{\text{4c,SD}}}a_{p}^{\dagger}a_{q}
\end{equation}
In Eq. (\ref{eq:4}), $A$ denotes the distinct atoms in the molecule,
$i$ referring to the occupied spinors for atom $A$, and $n_{i,A}$ denoting their respective occupation numbers. Consequently, the four-component DC Hamiltonian takes the form
\begin{equation}
\label{eq:5}
\hat{H^{\text{4c}}}=\sum_{pq}{h_{pq}^{\text{4c}}}a_{p}^{\dagger}a_{q}+\sum_{pq}{g_{pq}^{\text{4c,AMF}}}a_{p}^{\dagger}a_{q}+\frac{1}{4}\sum_{pqrs}{g_{pqrs}^{\text{4c,SF}}}a_{p}^{\dagger}a_{q}^{\dagger}a_{s}a_{r}
\end{equation}

The above Hamiltonian is now transformed into a two-component representation through the X2C decoupling scheme\cite{10.1063/1.473860}. First, the one-electron term $h^{\text{4c}}$ is written in matrix form using kinetically balanced basis functions.
\begin{equation}
\label{eq:6}
h^{\text{4c}}=\begin{pmatrix}
V & T\\
T & \frac{1}{4m^2c^2}W-T
\end{pmatrix}
\end{equation}

With $V$ representing the nuclear potential matrix, $T$ the kinetic energy matrix, and $W=(\boldsymbol{\sigma\cdot p})V(\boldsymbol{\sigma\cdot p})$ as the matrix form of the nuclear attraction for the small components. where $\boldsymbol{\sigma}$ refers to the Pauli spin matrices and $\boldsymbol{p}$ is the momentum operator. Thereafter, the $h^{\text{4c}}$ is fully block-diagonalized and transformed into the two-component representation by applying the following expression:
\begin{equation}
\label{eq:7}
h^{\text{X2C-1e}}=R^\dagger\left(V+X^\dagger T+TX+X^\dagger(\dfrac{1}{4m^2c^2}W-T)X\right)R
\end{equation}
with \textit{X} and \textit{R} as the X2C transformation matrices.

The term ($g_{pq}^{\text{4c,AMF}}$) in Eq. (\ref{eq:5}) is evaluated by solving the Dirac-Hartree-Fock equations for each distinct atom to obtain the atomic $X$ and $R$ matrices, which are then used to transform each AMF term into its two-component form ($g_{pq}^{\text{2c,AMF}}$).
In the absence of scalar two-electron picture-change (2e-pc) effects, the spin-free component in Eq. (\ref{eq:5}) reduces to nonrelativistic two-electron integrals. 
\begin{equation}
\label{eq:8}
g_{pqrs}^{\text{4c,SF}}\approx g_{pqrs}^{\text{NR}}
\end{equation}
Thus, the overall form of the two-component Hamiltonian in the X2CAMF scheme becomes
\begin{equation}
\label{eq:9}
\hat{H}^{\text{X2CAMF}}=\sum_{pq}{h_{pq}^{\text{X2C-1e}}}a_{p}^{\dagger}a_{q}+\sum_{pq}{g_{pq}^{\text{2c,AMF}}}a_{p}^{\dagger}a_{q}+\frac{1}{4}\sum_{pqrs}{g_{pqrs}^{\text{NR}}}a_{p}^{\dagger}a_{q}^{\dagger}a_{s}a_{r}
\end{equation}
The Hamiltonian above can be represented as an effective one-electron operator together with a nonrelativistic two-electron operator, as shown below:
\begin{equation}
\label{eq:10}
\hat{H}^{\text{X2CAMF}}=\sum_{pq}{h_{pq}^{\text{X2CAMF}}}a_{p}^{\dagger}a_{q}+\frac{1}{4}\sum_{pqrs}{g_{pqrs}^{\text{NR}}}a_{p}^{\dagger}a_{q}^{\dagger}a_{s}a_{r}
\end{equation}
with 
\begin{equation}
\label{eq:11}
h^{\text{X2CAMF}}=h^{\text{X2C-1e}}+g^{\text{2c,AMF}}
\end{equation}
The main advantage of employing the X2CAMF Hamiltonian is that it eliminates the need to compute molecular relativistic two-electron integrals, making it well-suited for relativistic calculations involving heavy-element systems.

\subsection{Relativistic Equation-of-Motion Coupled Cluster Method}
The exponential representation of the wave function serves as the basis for the CC theory\cite{Shavitt_Bartlett_2009}
\begin{equation}
\label{eq:12}
|\Psi_{\text{cc}}\rangle=e^{\hat{T}}|\Phi_{0}\rangle
\end{equation}
Here, $|\Phi_{0} \rangle$ represents the reference determinant and $\hat{T}$ is the cluster excitation operator. When the cluster operator is restricted to only single and double excitations, it leads to the commonly used CCSD (Coupled Cluster with Singles and Doubles) approximation.
The cluster amplitudes are determined by solving coupled sets of non-linear equations.
\begin{equation}
\label{eq:13}
\langle \Phi_{i}^{a}|\Bar{H}|\Phi_{0}\rangle=0
\end{equation}
\begin{equation}
\label{eq:14}
\langle \Phi_{ij}^{ab}|\Bar{H}|\Phi_{0}\rangle=0
\end{equation}
Where $ | \Phi_{i}^{a} \rangle$ and $ | \Phi_{ij}^{ab} \rangle$ as the singly and doubly excited determinants, respectively. The indices $i\hspace{0.1cm},j...$ and $a\hspace{0.1cm},b...$ refer to occupied and virtual spinors. The operator
$\Bar{H}=e^{-\hat{T}}\hat{H}^{\text{X2CAMF}}e^{\hat{T}}$ denotes the similarity transformed Hamiltonian, where $\hat{H}^{\text{X2CAMF}}$ is defined in Eq. (\ref{eq:10}).
The ground state energy within the CCSD approximation is given by the following expression
\begin{equation}
\label{eq:15}
\langle \Phi_{0}|\Bar{H}|\Phi_{0}\rangle=E
\end{equation}

In EOM-CC Theory, the target state $|\Psi_{k}\rangle$ is generated by applying a linear operator $\hat{R}_{k}$ on the CC ground state wavefunction
\begin{equation}
\label{eq:16}
|\Psi_{\text{k}}\rangle=\hat{R}_{k}|\Psi_{\text{cc}}\rangle
\end{equation}
The structure of the $\hat{R}_{k}$ operator is determined by the nature of the target state; in the case of the ionization potential variant of EOM-CC (IP-EOM-CC), it takes the form
\begin{equation}
\label{eq:17}
\hat{R}_{k}=\sum_{i}{r_{i}}a_{i}+\sum_{\substack{i < j \\ b}}{r_{ij}^{b}}a_{b}^{\dagger}a_{j}a_{i}+\cdots
\end{equation}
The difference in energy ($\omega_{k}=E_{k}-E_{0}$) between the target state and the ground state can be determined using the following equation
\begin{equation}
\label{eq:18}
[\Bar{H}, \hat{R}_{k}]|\Phi_{0}\rangle=\omega_{k}\hat{R}_{k}|\Phi_{0}\rangle
\end{equation}
Due to the non-Hermitian character of $\Bar{H}$, it also possesses a corresponding left eigenvector ($\langle \Phi_{0}|\hat{L}_{k}$), and it follows the biorthogonality
condition
\begin{equation}
\label{eq:19}
\langle \Phi_{0}|\hat{L}_{k}\hat{R}_{l}|\Phi_{0}\rangle=\delta_{kl}
\end{equation}
In the case of IP-EOM-CC, the $\hat{L}_{k}$ operator is expressed as
\begin{equation}
\label{eq:20}
\hat{L}_{k}=\sum_{i}{l^{i}}a_{i}^{\dagger}+\sum_{\substack{i < j \\ b}}{l_{b}^{ij}}a_{j}^{\dagger}a_{i}^{\dagger}a_{b}+\cdots
\end{equation}
Davidson's iterative diagonalization algorithm\cite{HIRAO1982246} is commonly used to solve the EOM-CC equations. As in the CCSD approach, the EOM-CC equations are typically truncated at the singles and doubles level, leading to the EOM-CCSD approximation. For the IP-EOM-CCSD method, the most computationally expensive step is the CCSD calculations as it scales as $O(N^{6})$ where \textit{N} is the total size of the basis.
Details of the EOM-CCSD theory are well-documented in the literature.\cite{Shavitt_Bartlett_2009}

\subsection{Cholesky decomposition}
Under Cholesky decomposition approximation\cite{https://doi.org/10.1002/qua.560120408}, we can approximate the two-electron repulsion integrals (ERIs) as a product of Cholesky vectors. 
\begin{equation}
\label{eq:21}
(\mu \nu| k\lambda)\approx \sum_{P}^{n_{\text{CD}}}L_{\mu \nu}^{P}L_{k\lambda}^{P},
\end{equation}
Here $\mu$, $\nu$, $k$, $\lambda$ represents atomic orbital (AO) indices, $L_{\mu \nu}^{P}$ is the Cholesky vector and $n_{\text{CD}}$ indicates the total number of Cholesky vectors. 
Both single-step\cite{https://doi.org/10.1002/qua.560120408,10.1063/1.1578621,AQUILANTE2007354,Pedersen2009} and two-step\cite{Aquilante2011,10.1063/1.5083802,doi:10.1021/acs.jpca.1c02317} algorithms have been proposed to perform the Cholesky decomposition of these integrals efficiently. In this work, we employ the conventional single-step algorithm, where Cholesky vectors are formed by an iterative procedure that continues until the largest diagonal element of the ERI matrix falls below the predefined Cholesky threshold ($\tau$). This threshold also serves to control the approximation error introduced by the decomposition.

After the formation of Cholesky vectors in the AO basis, we transform them into the MO basis 
\begin{equation}
\label{eq:22}
 L_{pq}^{P} = \sum_{\mu \nu}C_{\mu p}^{*}L_{\mu \nu}^{P}C_{\nu q},
\end{equation}
Upon transforming the Cholesky vectors to the molecular orbital (MO) basis, they can be used on-the-fly at any point to construct antisymmetrized two-electron integrals. In our implementation, we explicitly compute and store only those integrals involving two or fewer virtual indices, while the remaining integrals are constructed on the fly as needed.
\begin{equation}
\label{eq:23}
 \langle pq||rs \rangle = \sum_{P}^{n_{\text{CD}}}(L_{pr}^{P}L_{qs}^{P}-L_{ps}^{P}L_{qr}^{P}),
\end{equation}

\subsection{Frozen Natural Spinors}
Natural spinors are the relativistic analog of L\"owdin's natural orbitals\cite{PhysRev.97.1474}, which are obtained by diagonalizing the correlated one-body reduced density matrix (1-RDM) generated from a spin orbit-coupled wavefunction. The generation of MP2-based natural spinors can be accomplished by following these steps in order:
\begin{enumerate}
    \item Construct the virtual-virtual block of the 1-RDM ($D_{ab}$) using the MP2 method.
    \begin{equation}
    \label{eq:24}
    D_{ab}= \frac{1}{2}\sum_{cij}^{} {\frac{\langle ac||ij \rangle \hspace{0.1cm}\langle ij||bc \rangle}{\varepsilon_{ij}^{ac} \hspace{0.2cm}\varepsilon_{ij}^{bc}}}
    \end{equation}
    \item Diagonalization of $D_{ab}$ to get eigenvectors (\textit{V}) as virtual natural spinors and the corresponding eigenvalues (\textit{n}) as occupation numbers.
    \begin{equation}
    \label{eq:25}
    D_{ab}V=Vn
    \end{equation}
    \item Conversion of the virtual-virtual block of the Fock matrix ($F_{vv}$) into the truncated basis of natural spinors ($F^{\text{NS}}_{vv}$) by setting a predetermined threshold or cutoff for the
    occupation numbers
    \begin{equation}
    \label{eq:26}
    F^{\text{NS}}_{vv}=\Tilde{V}^{\dagger}F_{vv}\Tilde{V}
    \end{equation}
    with $\Tilde{V}$ as virtual natural spinors in a truncated basis.
    \item Diagonalization of the $F_{vv}^{\text{NS}}$ block to get the semi-canonical virtual natural spinors ($\tilde{Z}$) and their associated orbital energies ($\epsilon$).
    \begin{equation}
    \label{eq:27}
    F_{vv}^{\text{NS}}\tilde{Z}=\tilde{Z}\epsilon
    \end{equation}
    \item Creation of the transformation matrix ($U$) for the conversion of the canonical virtual spinor space to the semi-canonical natural virtual spinor space.
    \begin{equation}
    \label{eq:28}
    U=\tilde{Z}\tilde{V}
    \end{equation}
\end{enumerate}
The set of spinors created is called the "Frozen Natural Spinor (FNS)", as the occupied spinors were kept intact in their canonical form, and we only transform the virtual spinors into semi-canonical natural virtual spinors.

\section{Implementation and Computational Details}
The FNS-CD-X2CAMF-IP-EOM-CCSD method has been implemented within the development version of
BAGH\cite{dutta2023bagh}. The X2CAMF-HF computations are carried out using the \texttt{socutils} package\cite{socutils}, which is interfaced with BAGH.
The FNS- and CD-based X2CAMF-IP-EOM-CCSD implementation builds upon the existing ground-state FNS-CD-X2CAMF-CCSD framework\cite{doi:10.1021/acs.jctc.5c00199}. It utilizes the same Cholesky-decomposed and FNS-based integral infrastructure up to the CCSD level, which includes the generation of Cholesky vectors in the AO basis, partial transformation to the MO basis for LOO and LOV-type Cholesky vectors, formation of frozen natural spinors via 1-RDM diagonalization, and the construction of transformed Cholesky vectors and antisymmetrized 0-2 particle (external) integrals in the FNS basis.
In the IP-EOM stage, one would require the construction of  $\Bar{H}$ intermediate up to two virtual intermediates, which are constructed only once from the converged coupled cluster amplitudes and stored.  
To assess the performance of the FNS-CD-X2CAMF-IP-EOM-CCSD method, we carried out a comprehensive analysis on two molecular test sets derived from the SOC-81 dataset of Scherpelz et al\cite{doi:10.1021/acs.jctc.6b00114}: a subset of 18 iodine-containing molecules, and a subset of 74 heavier-element-containing molecules selected from the SOC-81 dataset. The latter was also employed by Zgid and coworkers\cite{doi:10.1021/acs.jctc.4c00075} in their fully self-consistent GW calculations for vertical ionization potentials.
We have also simulated the photoelectron spectra of cadmium halides (CdX$_{2}$, X = Cl, Br, I) employing the FNS-CD-X2CAMF-IP-EOM-CCSD method and compared the results with experimental spectra. The experimental spectra were digitized from the data provided in the work of Schweitzer and coworkers\cite{BOGGESS1973467} via the WebPlotDigitizer\cite{WebPlotDigitizer} tool.
The molecular geometries of all test case molecules were obtained from Ref \cite{doi:10.1021/acs.jctc.6b00114}, and experimental vertical ionization potentials were taken from the NIST WebBook\cite{nist_webbook_2025}. All calculations used the dyall.v4z basis set, and the frozen-core approximation was applied throughout unless otherwise stated.

\section{Results and Discussion}
\subsection{Convergence with Respect to the Threshold}
To determine the optimal threshold that balances computational efficiency and accuracy for ionization potential calculations, we have considered three truncation thresholds: LOOSEFNS (FNS threshold: $10^{-4}$, and CD threshold: $10^{-3}$), NORMALFNS (FNS threshold: $10^{-4.5}$, and CD threshold: $10^{-4}$), and TIGHTFNS (FNS threshold: $10^{-5}$, and CD threshold: $10^{-5}$), following the approach used in our previous work on ground-state FNS-CD-X2CAMF-CCSD implementation\cite{doi:10.1021/acs.jctc.5c00199}. In that study, among the three thresholds, NORMALFNS was found to offer the best balance between computational cost and accuracy for achieving converged ground-state energies and properties. However, for ionization potential calculations in the present work, it is necessary to reassess the performance of all three thresholds. Therefore, we performed FNS-CD-X2CAMF-IP-EOM-CCSD calculations on a subset of 18 iodine-containing molecules to compute the first valence ionization potentials at each threshold. For each case, the mean error (ME), mean absolute error (MAE), and standard error (SE) with respect to the experimental values were evaluated. Table \ref{table:1} presents the ME, MAE, and SE of vertical ionization energies computed using the FNS-CD-X2CAMF-IP-EOM-CCSD method with respect to the experimental values under different thresholds. The absolute values of the ionization energies for the subset are provided in Table S1 of the supporting information. The results show that the error magnitudes are very similar across all three settings, with ME values within $\pm$0.008 eV. The MAE and SE values among the three thresholds also differ by no more than 0.004 eV, indicating that the accuracy remains largely unaffected by the choice of threshold. Given the minimal variation in accuracy, the use of a LOOSEFNS threshold appears justified, as it can significantly reduce computational cost while maintaining reliable results. Thus, LOOSEFNS was adopted for all subsequent calculations because of its favorable compromise between computational cost and precision.
\begin{table}[H]
\centering
\caption{Convergence of the ME, MAE, and SE of vertical ionization energies computed using the FNS-CD-X2CAMF-IP-EOM-CCSD method with respect to the experimental values under different thresholds for a set of 18 iodine-containing molecules\textsuperscript{a}}
\begin{threeparttable}
\begin{tabular}{c c c c}
\hline \hline \\
  & \multicolumn{3}{c}{Vertical ionization energy (eV)} \\
 \cline{2-4} \\
 Threshold & ME & MAE & SE   \\
 \hline \\
 LOOSEFNS  & 0.005 & 0.068 & 0.081 \\
 NORMALFNS & 0.008 & 0.066 & 0.078 \\
 TIGHTFNS  & 0.003 & 0.065 & 0.077 \\
\hline \hline 
\end{tabular}
\begin{tablenotes}
\item[a] The basis set used is dyall.v4z
\end{tablenotes}
\end{threeparttable}
\label{table:1}
\end{table}

\subsection{Benchmark Calculations for SOC-81 Subset}
\begin{figure}[H]
    \begin{center}
    \includegraphics[width=0.7\textwidth]{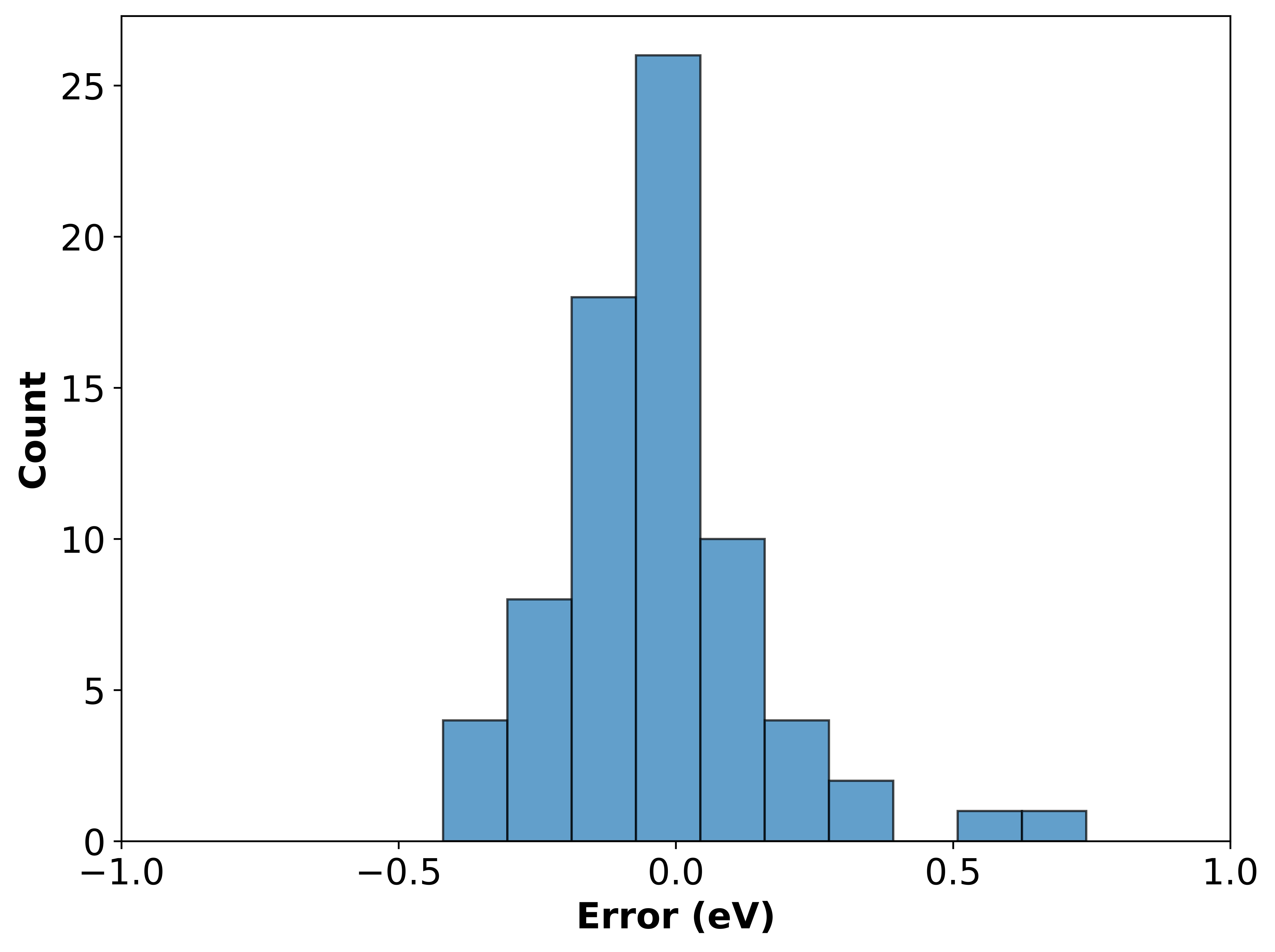}
    \caption{Distribution of errors in vertical ionization energy using the FNS-CD-X2CAMF-IP-EOM-CCSD method relative to experimental values for the selected subset of molecules.}
    \label{fig:my_label1}
    \end{center}
\end{figure}

Having established LOOSEFNS as the optimal truncation threshold, we now adopt this threshold setting as the basis for benchmarking the new set of calculations, which involves a subset of 74 heavier-element-containing molecules selected from the SOC-81 dataset. For this subset, we carried out FNS-CD-X2CAMF-IP-EOM-CCSD calculations and evaluated the performance of the method by calculating the ME, MAE, and SE relative to experimental ionization energies. Table \ref{table:2} summarizes the ME, MAE, and SE for the first vertical ionization energies obtained using the FNS-CD-X2CAMF-IP-EOM-CCSD method. The corresponding absolute ionization energy values are provided in Table S2. Previously reported values from Zgid and co-workers\cite{doi:10.1021/acs.jctc.4c00075} using the fully self-consistent GW method for the same subset are also included for comparison. 
From Table \ref{table:2}, it can be seen that in terms of MAE, FNS-CD-X2CAMF-IP-EOM-CCSD achieves slightly reduced errors, though the differences are modest relative to X2C(G0W0@PBE0) and X2C(scGW) methods. The SEs are also comparable across methods, indicating similar levels of consistency.  The ME values suggest that both X2C(G0W0@PBE0) and FNS-CD-X2CAMF-IP-EOM-CCSD yield results that are close to the reference data, with minimal deviation. Figure \ref{fig:my_label1} shows the distribution of errors in ionization energy predictions using the FNS-CD-X2CAMF-IP-EOM-CCSD method relative to experimental values for the selected subset, providing a visual overview of the method's overall accuracy and error behavior. These results suggest that the FNS-CD-X2CAMF-IP-EOM-CCSD method offers a reliable, accurate, and computationally efficient alternative for evaluating vertical ionization energies, especially in systems containing heavy elements.

\begin{table}[t]
\centering
\caption{ME, MAE, and SE of vertical ionization energies computed using the FNS-CD-X2CAMF-IP-EOM-CCSD method with respect to the experimental values and their comparison with previously reported values}
\begin{threeparttable}
\begin{tabular}{ c c c c}
\hline \hline \\
  & \multicolumn{3}{c}{Vertical ionization energy (eV)} \\
 \cline{2-4} \\
 Method & ME  & MAE & SE   \\
 \hline \\
 X2C(G0W0@PBE)\cite{doi:10.1021/acs.jctc.4c00075}  & -0.31 & 0.32 & 0.23  \\
 X2C(G0W0@PBE0)\cite{doi:10.1021/acs.jctc.4c00075} & 0.00 & 0.14 & 0.20 \\
 X2C(scGW)\cite{doi:10.1021/acs.jctc.4c00075}  & -0.17 & 0.21 & 0.17 \\
 FNS-CD-X2CAMF-IP-EOM-CCSD & -0.02 & 0.13 & 0.18 \\
\hline \hline 
\end{tabular}
\end{threeparttable}
\label{table:2}
\end{table}

\subsection{Photoelectron Spectra of Cadmium Halides.}
\begin{figure}[H]
    \begin{center}
    \includegraphics[width=1.0\textwidth]{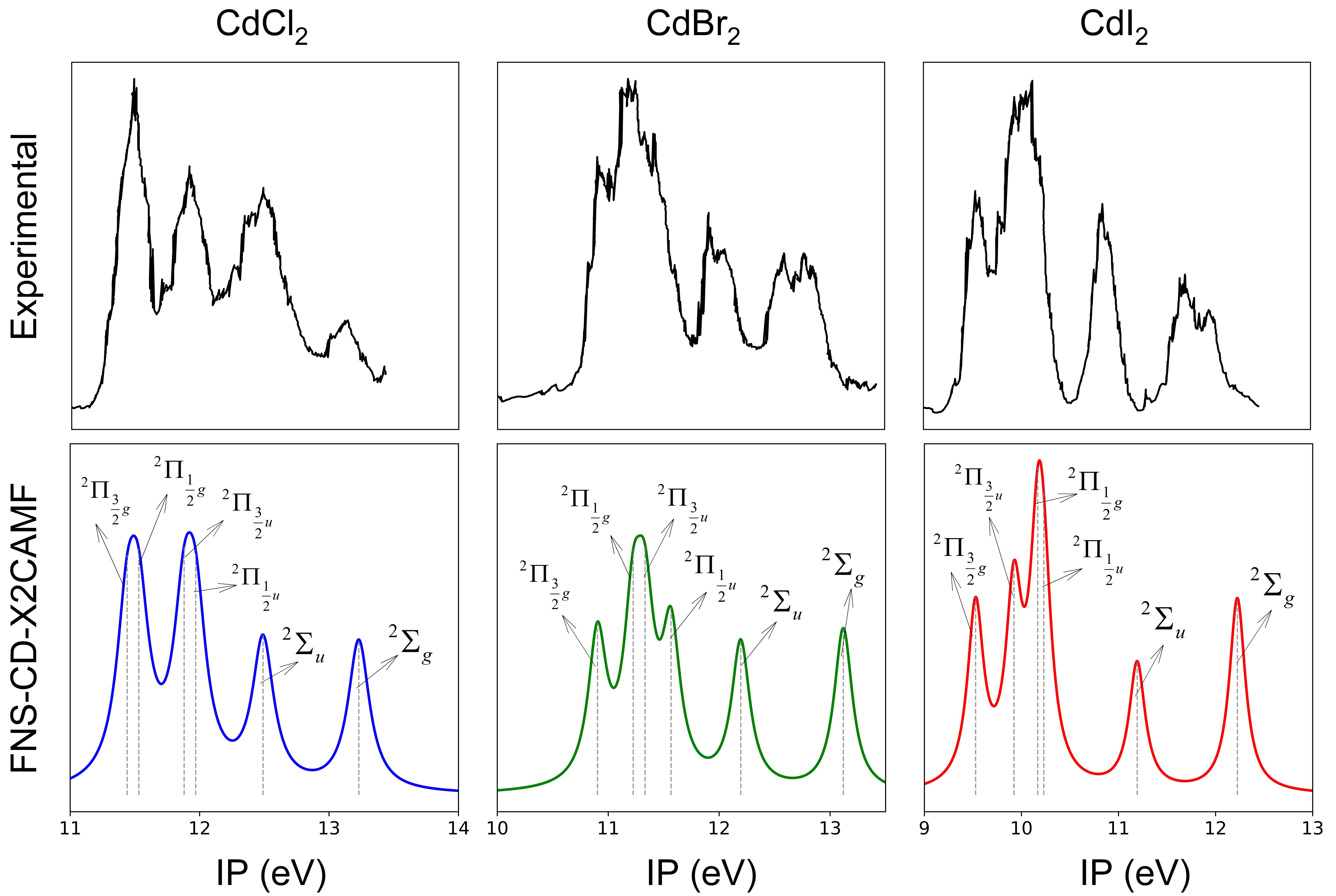}
    \caption{Comparison of experimental photoelectron spectra with the spectra obtained using FNS-CD-X2CAMF-IP-EOM-CCSD method for Cadmium halides.}
    \label{fig:my_label2}
    \end{center}
\end{figure}

Transition probabilities obtained from IP calculations can be utilized to simulate photoelectron spectra. Cadmium halides (CdX$_{2}$, X=Cl, Br, I) have often served as benchmark systems for testing the accuracy of such calculations by comparing simulated and experimental photoelectron spectra\cite{doi:10.1021/acs.jctc.4c00075,doi:10.1021/acs.jctc.4c01762}. In this section, we aim to further assess the performance of the FNS-CD-X2CAMF-IP-EOM-CCSD method by generating and analyzing the photoelectron spectra of cadmium halides. Figure \ref{fig:my_label2} displays the photoelectron spectra of the CdX$_{2}$ series obtained using the FNS-CD-X2CAMF-IP-EOM-CCSD method, along with a comparison to the experimental spectra\cite{BOGGESS1973467}. For direct comparison with experimental spectrum, a uniform energy shift (-0.02 eV for CdCl$_{2}$, +0.2 eV for CdBr$_{2}$, and -0.12 eV for CdI$_{2}$) was applied to the computed spectra to match the position of their lowest-energy peaks with those in the experimental data. The figure clearly shows that the FNS-CD-X2CAMF-IP-EOM-CCSD method effectively replicates the overall peak pattern observed in the experimental spectra. The influence of spin-orbit coupling becomes increasingly pronounced from Cl to I, with CdI$_{2}$ exhibiting the strongest splitting in the series. In this case, the $^2\Pi_{\frac{1}{2}\mathrm{\textit{g}}}$ and $^2\Pi_{\frac{3}{2}\mathrm{\textit{u}}}$ states are reordered, a feature that should be reflected in the spectra. The computed spectra capture this reordering, highlighting the capability of the method to account for spin-orbit relativistic effects. In addition to being consistent with experimental data, the computed spectra also align with results from other recent theoretical studies found in the literature\cite{doi:10.1021/acs.jctc.4c00075,doi:10.1021/acs.jctc.4c01762}.

\subsection{Comparison with the Four-Component FNS results}
To validate the accuracy of the FNS-CD-X2CAMF-IP-EOM-CCSD results, we carried out a direct comparison with vertical ionization potentials computed using the four-component FNS-IP-EOM-CCSD method for the hydrogen halide series (HX, X=F, Cl, Br, I, At). The four-component IP values are taken from our earlier study\cite{10.1063/5.0125868}. For H, F, and Cl, an uncontracted aug-cc-pVQZ basis set was employed, while the dyall.ae4z basis set was used for Br, I, and At. The frozen core approximation was applied only to HI and HAt. The FNS truncation threshold in the previous study\cite{10.1063/5.0125868} was set to $10^{-5}$. Therefore, TTIGHTFNS (FNS threshold: $10^{-5}$, and CD threshold: $10^{-5}$) has been used for the hydrogen halide series to maintain consistency with the previous report\cite{10.1063/5.0125868}. Table \ref{table:3} compares vertical ionization energies for the first three states of the HX series, obtained using the four-component FNS-IP-EOM-CCSD, two-component spin-free X2C in its one-electron variant (SFX2C1e) based FNS-IP-EOM-CCSD, and two-component FNS-CD-X2CAMF-IP-EOM-CCSD approaches. From Table \ref{table:3}, it is clear that both the four-component and CD-X2CAMF methods produce nearly identical IP values, with differences typically no greater than 0.001 eV. Within the four-component framework, it has been shown that spin-orbit coupling significantly impacts the accuracy of IP values computed using the IP-EOM-CCSD method. Relying solely on the spin-free X2C Hamiltonian proves insufficient, particularly for systems containing heavier elements. The X2CAMF, however, maintains its accuracy even for heavier halides like HI and HAt, effectively capturing both scalar and spin-orbit relativistic effects. Furthermore, both the four-component and CD-X2CAMF approaches show strong consistency with experimental ionization potentials wherever data are available, with deviations typically within 0.05 eV. This further supports the reliability of the X2CAMF approximation, demonstrating its ability to not only replicate theoretical results but also closely match experimental findings. Overall, this comparison highlights that the two-component FNS-CD-X2CAMF-IP-EOM-CCSD method provides a robust, efficient, and accurate alternative to more computationally expensive four-component calculations, making it particularly well-suited for predicting ionization energies in relativistic systems with heavy atoms.

\begin{table}[H]
\centering
\caption{Comparison of ionization potential values calculated using the four-component FNS-IP-EOM-CCSD method\cite{10.1063/5.0125868}, FNS-SFX2C1e-IP-EOM-CCSD method\cite{10.1063/5.0125868}, and the FNS-CD-X2CAMF-IP-EOM-CCSD method. The basis used is uncontracted aug-cc-pVQZ for H, F, and Cl, and the dyall.ae4z for Br, I, and At.}
\begin{threeparttable}
\begin{tabular}{ c c c c c c }
\hline \hline
 && \multicolumn{3}{c}{Vertical Ionization Potential (eV)}  \\
\cline{3-5}
Molecule & Ionization State & 4c\tnote{a} & SFX2C1e\tnote{b}&X2CAMF\tnote{c} & Experimental \\
\hline
HF  & 5 $\Pi$ & 16.106 & 16.151 &16.106 &16.120\cite{10.1063/1.431263} \\
    & 4 $\Pi$ & 16.145 & 16.151  &16.145 &-- \\
   & 3 $\Sigma$ & 20.042 &20.063 & 20.042 & 19.890\cite{10.1063/1.431263}\\
\hline
HCl & 9 $\Pi$ & 12.768 &12.826  & 12.768 &12.745\cite{YENCHA1998109} \\
 & 8 $\Pi$ &12.852 &12.826 & 12.852 &	12.830\cite{YENCHA1998109} \\
  & 7 $\Sigma$ & 16.791 & 16.804 & 16.791 &-- \\
\hline
HBr  & 18 $\Pi$ & 11.689 & 11.861 & 11.689 &11.645\cite{ADAM1992185} \\
 & 17 $\Pi$ & 12.025&  11.861 & 12.026&11.980\cite{ADAM1992185} \\
  & 16 $\Sigma$ & 15.811 & 15.802&15.810 &15.650\cite{ADAM1992185} \\
\hline
HI  & 27 $\Pi$ & 10.425 & 10.759 &10.425 &10.388\cite{CORMACK1997175} \\
 & 26 $\Pi$ &11.091 & 10.759&11.091 & 11.047\cite{CORMACK1997175} \\
  & 25 $\Sigma$ & 14.461 &14.423 & 14.461 &-- \\
\hline
HAt  & 43 $\Pi$ & 9.290&10.214 & 9.291 & --\\
 & 42 $\Pi$ & 11.035& 10.214& 11.034 &-- \\
  & 41 $\Sigma$ & 14.240 &13.841 & 14.240 &-- \\
\hline \hline
\end{tabular}
\begin{tablenotes}
\item[a] Four-Component FNS-IP-EOM-CCSD. Taken from \cite{10.1063/5.0125868}
\item[b] FNS-SFX2C1e-IP-EOM-CCSD. Taken from \cite{10.1063/5.0125868}
\item[c] FNS-CD-X2CAMF-IP-EOM-CCSD.
\end{tablenotes}
\end{threeparttable}
\label{table:3}
\end{table}

\subsection{Application to Medium-Sized Complex}
\begin{figure}[H]
    \begin{center}
    \includegraphics[width=0.5\textwidth]{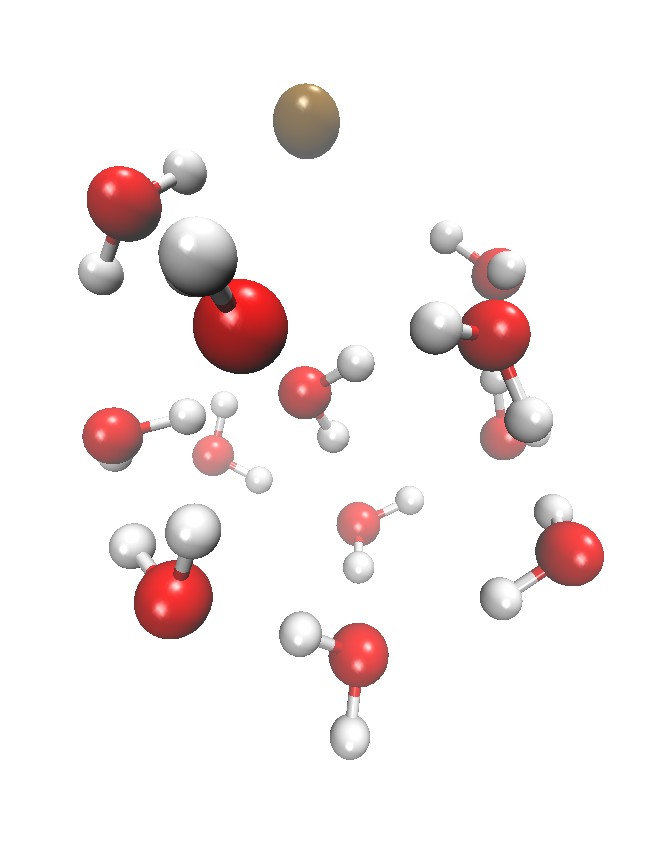}
    \caption{Molecular structure for the [I(H$_{2}$O)$_{12}$]$^{-}$ complex.}
    \label{fig:my_label3}
    \end{center}
\end{figure}

To demonstrate the applicability of the FNS-CD-X2CAMF-IP-EOM-CCSD method to medium-sized molecular systems, we applied it to the [I(H$_{2}$O)$_{12}$]$^{-}$ complex to compute its vertical ionization energies. Optimized geometry of the complex was obtained at the DFT level using the B3LYP functional with the inclusion of scalar relativistic effects via the zero-order regular approximation (ZORA). The ORCA-6.0 software package\cite{ORCA} has been used for geometry optimization. The SARC-ZORA-TZVP basis set was used for iodine atoms, and the def2-TZVP basis set was used for the rest of the atoms. Figure \ref{fig:my_label3} displays the optimized molecular structure of the complex. For the FNS-CD-X2CAMF-IP-EOM-CCSD calculations, an uncontracted aug-cc-pVDZ basis set was employed for the H and O atoms, while the s-aug-dyall.v4z basis set was used for the I atom. The [I(H$_{2}$O)$_{12}$]$^{-}$ complex features a basis set of dimension 1872, comprising 174 occupied and 1698 virtual spinors. In the correlation treatment, core electrons were frozen, and a LOOSEFNS truncation threshold was applied, reducing the system to 114 occupied and 648 virtual spinors. At this threshold, a total of 2574 Cholesky vectors were generated. The calculations were performed sequentially on a dedicated workstation featuring dual Intel(R) Xeon(R) Gold 5315Y processors running at 3.20 GHz and 2.0 TB of RAM. The Cholesky vector formation in the AO basis took 23 minutes and 52 seconds. The time required for the construction of two-electron integrals on the basis of the FNS was 1 hour, 11 minutes, and 35 seconds. The CCSD calculation took 3 days, 42 minutes, while the EOM calculations required 1 day, 11 hours, and 13 minutes. In total, the time required to compute the first four roots using the FNS-CD-X2CAMF-IP-EOM-CCSD method for the [I(H$_{2}$O)$_{12}$]$^{-}$ complex was 4 days, 19 hours, and 5 minutes. The first vertical ionization energy was found to be 4.30 eV. The vertical ionization potential of isolated iodine at FNS-CD-X2CAMF-IP-EOM-CCSD\textbackslash s-aug-dyall.v4z level of theory is found to be 3.07 eV, which is in excellent agreement with the experimental value of 3.06 eV\cite{10.1063/1.444805,10.1063/1.461172}. The micro-solvation with twelve water molecules causes a blue shift of 1.23 eV, which is consistent with the trend reported by Gomes and coworkers \cite{PhysRevLett.121.266001}.

\section{Conclusions}
We present an efficient implementation of the relativistic IP-EOM-CCSD method, based on the FNS- and CD-based X2CAMF Hamiltonian, aimed at significantly reducing computational cost while maintaining similar accuracy as that of the four-component Dirac-Coulomb Hamiltonian. The present formulation avoids the construction and storage of integrals and intermediates involving three or four virtual orbital indices and reduces floating-point operations, making it well-suited for molecules containing heavy atoms. Benchmark calculations on a set of 18 iodine-containing molecules demonstrate that using a LOOSEFNS (FNS threshold: $10^{-4}$, and CD threshold: $10^{-3}$) threshold yields a favorable balance between computational efficiency and chemical accuracy. The method also reliably reproduces experimental ionization energies for 74 heavy-element-containing molecules from the SOC-81 dataset with an MAE of 0.13 eV only, and accurately captures photoelectron spectra for cadmium halides. Furthermore, its practical applicability is showcased through IP calculations on the [I(H$_{2}$O)$_{12}$]$^{-}$ complex, demonstrating the method's suitability for routine, accurate relativistic IP-EOM-CCSD calculations on medium-sized systems. The inclusion of triples correction within the FNS-CD-X2CAMF-IP-EOM-CC framework can further improve the accuracy of the calculations. Work is in progress towards that direction. 

\begin{acknowledgement}
The authors acknowledge the support from IIT Bombay, CRG, and Matrix project of DST-SERB, CSIR-India, DST-Inspire Faculty Fellowship, Prime Minister's Research Fellowship, ISRO, for financial support, IIT Bombay supercomputational facility, and C-DAC Supercomputing resources (PARAM Yuva-II, PARAM Bramha) for computational time. AKD acknowledges the research fellowship funded by the EU NextGenerationEU through the Recovery and Resilience Plan for Slovakia under project No. 09I03-03-V04-00117. The authors acknowledge Xubo Wang (Department of Chemistry, Johns Hopkins University) and Chaoqun Zhang (Department of Chemistry, Yale University) for their valuable feedback and discussions during the preparation of this manuscript.
\end{acknowledgement}

\begin{suppinfo}
The following file is available free of charge.
\begin{itemize}
  \item SI: The absolute values of the ionization energies for a subset of 18 iodine-containing molecules, and for a subset of 74 heavier-element-containing molecules selected from the SOC-81 dataset.
\end{itemize}
\end{suppinfo}

\bibliography{main}

\end{document}